\documentstyle[aps,epsfig,twocolumn]{revtex}    
\draft    
    
\begin{document}

\title{Bose-Einstein    
condensates in  1D optical lattices:    
compressibility, Bloch bands and elementary excitations}  
\author{$^1$M. Kr\"amer, $^{1,2}$C. Menotti, $^{1,3}$L. Pitaevskii     
and $^1$S. Stringari}    
\address{$^1$ Dipartimento di Fisica, Universit\`{a} di Trento and BEC-INFM,  
I-38050 Povo, Italy}    
\address{$^2$ Dipartimento di Matematica e Fisica,    
Universit\`{a} Cattolica del Sacro Cuore,  
I-25121 Brescia, Italy}    
\address{$^3$ Kapitza Institute for Physical Problems, 117334 Moscow, Russia}  
    
\date{\today}    
    
\maketitle    
    
\begin{abstract}  
\noindent  
We discuss the Bloch-state solutions of the stationary Gross-Pitaevskii 
equation and of the Bogoliubov equations for a Bose-Einstein condensate in the presence of a one-dimensional optical lattice. 
The results for the compressibility, effective mass and velocity of sound are analysed as a function of the lattice depth and of the strength of the two-body interaction. 
The band structure of the spectrum of elementary excitations is compared with the one exhibited by the stationary solutions (``Bloch bands''). 
Moreover, the numerical calculations are compared with the analytic predictions of the tight binding approximation. 
We also discuss the role of quantum fluctuations and show that the condensate 
 exhibits 3D, 2D or 1D features depending on the lattice depth and on the number of particles occupying each potential well.
We finally show how, using a local density approximation, our results can be applied to study 
the behaviour of the gas in the presence of harmonic trapping. 
\\
\end{abstract}

\section{Introduction}    
    
Cold atoms in optical lattices exhibit phenomena typical 
of solid state physics like the formation of energy bands, Bloch oscillations,
and Josephson effects. Many of these phenomena have been already the
object of experimental 
and 
theoretical 
investigation in Bose-Einstein condensates.  For deep potential wells
further important effects take place like the transition from the
superfluid to the Mott insulator phase \cite{jaksch,greiner}.
     
The purpose of this paper is to study some structural properties of
interacting Bose-Einstein condensed dilute gases at $T=0$ in the presence of 1D
periodic potentials generated by laser fields (1D optical lattices).
Unless the confinement in the radial direction is very tight,
 the transition to the insulator phase in 1D optical lattices is expected to take place only for extremely deep potential wells. 
There is consequently a large range of potential depths where
the gas can be described as a fully coherent system,
in the framework of 
the mean field Gross-Pitaevskii approach to the
order parameter. 
 
We will explicitly discuss the change in the behaviour of the system
as a function of the optical potential depth, ranging from the
uniform gas (absence of optical lattice) to the opposite regime of
deep wells separated by high barrieres (tight
binding limit).
Special emphasis will be given to the role of two-body interactions,
whose effects will be adressed by varying the average density $n$.
 
An important feature produced by the periodic potential is the
occurrence of a typical band structure in the energy spectra.  In
this paper we will discuss different manifestations of such a band
structure, including:
  
- the energy per particle $\varepsilon_j(k)$ of stationary Bloch-wave 
  configurations consisting in the motion of the whole condensate 
and carrying a current constant in time and uniform in space (``Bloch bands'').  This energy
  is naturally parametrized as a function of the quasimomentum 
$k$ which, together with the band index $j$ ($j=1,2,...$), is the
  proper quantum number of these states;
  
- the chemical potential
\begin{equation}
\mu_j (k)= {\partial [n \varepsilon_j (k) ] \over \partial n}\,, 
\label{mu_l}
\end{equation}
of the same stationary configurations,
where $n$ is the average density of the sample.  The chemical
potential plays a crucial role in the determination of the equation of
state and emerges as a natural output of the solution of the
Gross-Pitaevskii equation (see Eq.(\ref{galileo}) below);
  
- the spectrum $\hbar
\omega_j(q)$ of the elementary excitations (``Bogoliubov bands'') 
carrying quasi-momentum $q$. 
The elementary excitations are small perturbations of the system and in general can be calculated with respect to each stationary configuration of quasimomentum $k$. 
In this paper we will limit the discussion to the elementary excitations built on top of the groundstate configuration ($k=0$). 
They can be determined by solving the
linearized Gross-Pitaevskii equations (see
Eqs.(\ref{bog_u},\ref{bog_v})).
  
The three band spectra $\varepsilon_j$, $\mu_j$ and $\hbar
\omega_j$ represent different physical quantities and have the same dependence on quasimomentum 
only in the absence of two body interactions.  Due to
the periodicity of the problem the quasi-momentum can be restricted to the
first Brillouin zone. Still it is often convenient to consider all
values of quasi-momentum to emphasize the periodicity of the
energy spectra in quasi-momentum space.
  
In Sect.\ref{section2}, we discuss the equation
of state and the Bloch bands, assuming uniform confinement in the
radial direction.
Special emphasis is given to the behaviour of the compressibility
and the effective mass.  The compressibility $\kappa$ is
defined by the thermodynamic relation
     
\begin{eqnarray}    
\kappa^{-1 } = n {\partial \mu \over \partial n} \,,
\label{kappa}    
\end{eqnarray}    
where $\mu$ is the chemical potential relative to the groundstate
solution of the Gross-Pitaevskii equation ($\mu\equiv
\mu_{j=1}(k=0)$).  For systems interacting with repulsive forces, the
optical trapping reduces the compressibility of the system since the effect of repulsion  
is enhanced by the squeezing of the condensate wavefunction in each well.  
The effective mass is defined through the
curvature of the lowest ($j=1$) energy band
    
\begin{eqnarray}    
{1 \over m^*(k)} =   
 {\partial^2 \varepsilon\over \partial k^2}\,.
\label{curvature}    
\end{eqnarray}    
Here and in the following, we omit the band index $j$ when
we refer to the lowest band ($j=1$).
The current flowing along the direction
of the optical lattice, is fixed by the relation
     
\begin{eqnarray}    
I (k) = n {\partial \varepsilon\over \partial k}\,,      
\label{current}    
\end{eqnarray}    
and in the long wavelength limit $k \to 0$, one finds $I \to n k
/m^*$, where we have used the notation $m^*\equiv m^*(k=0)$.  
Equivalently, one has
\begin{eqnarray}    
\varepsilon(k)
\raisebox{0.1cm}{$  
\;\;\;\; \longrightarrow \;\;\;$}  
\hspace*{-1.15cm}   
\raisebox{-0.15cm}{$\; k \to 0 $} \;\;\;  
\varepsilon(k=0)+{k^2\over2m^*}\,.
\label{epsilonlowk}
\end{eqnarray}    
For
small intensities of the laser field the effective mass $m^*$
approaches the bare value $m$ .  Instead, for large intensities the
effective mass is inversely proportional to the tunneling rate through
the barrier separating neighbouring potential wells and is strongly
enhanced with respect to the bare value.

Based on the knowledge of the compressibility and of the   
effective mass, one can calculate the sound velocity according to the   
thermodynamic relation 
 
\begin{eqnarray}    
c = 
{1\over\sqrt{\kappa m^*} }\,.
\label{sound_vel}    
\end{eqnarray}    
This quantity fixes the slope of the lowest Bogoliubov band 
at low quasi momenta. 
    
In Sect.\ref{elementary-excitations}, we discuss the behaviour of the
elementary excitations.  The excitation spectrum (Bogoliubov
spectrum) is obtained from the solution of the linearized time-dependent
Gross-Pitaevskii equation and develops energy bands $\hbar\omega_j(q)$ periodic in 
quasi-momentum space.  
From the same solutions one can calculate the excitation strengths $Z_j(p)$ relative to the density operator and hence the dynamic structure factor $S(p,\omega)$ \cite{chiara}. 
Here, $p$ and $\hbar\omega$ are the momentum and the energy transferred by a weak external probe. 
Differently from the Bogoliubov energies $\hbar \omega_j$, 
the strengths $Z_j(p)$, and 
$S(p,\omega)$, are not periodic functions of $p$, putting in clear
evidence the difference between {\it momentum} and {\it
quasi-momentum}.  
    
We will often compare our numerical results with the predictions of the
so called tight binding approximation, which is reached when the
intensity of the laser field generating the optical lattice is so high
that only the overlap between the wavefunctions of nearest-neighbour
condensates plays a role.  In the tight binding limit, the lowest Bloch
and Bogoliubov bands take analytic forms that will be discussed explicitly.
    
In Sect.\ref{fluct_depl}, we study the effect of the lattice on the
quantum depletion of the condensate and comment on the validity of 
Gross-Pitaevskii and Bogoliubov theory. Depending on
the parameters of the problem, we distinguish between different
configurations which show 3D, 2D or 1D features.

Finally, in Sect.\ref{appl_to_ho} we  include the presence of an
additional harmonic trapping. This is achieved by employing a local
density approximation to treat the inhomogeneity of the density
profile due to the harmonic confinement.  Important applications
concern the frequencies of collective oscillations.

\section{Compressibility  and effective mass}    
\label{section2}    
    
The aim of this section is to calculate the compressibility and the   
effective mass as a function of the  average density of the gas and of   
the depth of the optical potential.     
Using these quantities, we will also calculate the velocity of sound.    
 
We consider the following geometry: along the $z$     
direction the atoms feel the periodic potential created by two  
counterpropagating laser fields, while the system is uniform in the   
transverse direction.    
The important length, momentum and energy scales of the problem are:
the lattice spacing $d={ \pi / k_{\rm opt} }$ related to the
wavevector $k_{\rm opt}$ of the laser field in the lattice direction,
the Bragg momentum $q_B=\hbar k_{\rm opt}={\hbar \pi / d }$ which identifies the edge of
the Brillouin zone and the recoil energy $E_R={\hbar^2 \pi^2 /
2md^2}$, corresponding to the energy gained by an atom at rest by
absorbing a lattice photon.
    
The  trapping potential generated by the optical field can be simply     
written in the form    
    
\begin{eqnarray}    
V(z)= s\; E_R   \;{\rm sin}^2\left( {\pi z \over d} \right),    
\label{Vext}    
\end{eqnarray}    
where $s$ is a dimensionless parameter which denotes the lattice depth      
in units of the recoil energy $E_R$.    
    
The parameter characterizing the role of interactions in the system is
$g n$, defined as the two-body coupling constant $g=4 \pi \hbar^2 a
/m$ times the 3D {\it average} density $n$. Here $a$ is the s-wave
scattering length which will be always assumed to be positive.
Typical values of the ratio $gn/E_R$ used in recent experiments 
\cite{denschlag,cataliotti,morsch1,greiner1} range from $0.02$ to $1$.

Since the density is uniform in the transverse direction, the
transverse degrees of freedom decouple from the axial ones and the stationary 
Gross-Pitaevskii equation can be reduced to a 1D equation of the form
\begin{eqnarray}    
\left[-{\hbar^2 \over 2m}\;{\partial^2\over\partial z}   
+s\;E_R \; {\rm sin}^2\left({ \pi z \over d  }\right)    
+{g n d  }|\varphi(z)|^2    
\right]\varphi(z)= \nonumber \\  
= {\mu }\varphi(z)\,,    
\label{gpe}    
\end{eqnarray}    
where the order parameter $\varphi$ is normalized according to 
$\int_{-d/2}^{d/2}|\varphi(z)|^2 dz =1$.

In spite of its non-linearity, Eq.(\ref{gpe}) permits solutions in the
form of Bloch waves
\begin{eqnarray}    
\varphi_{jk}(z)=e^{ikz/\hbar} \tilde{\varphi}_{jk}(z)\,,
\label{bloch-ansatz}    
\end{eqnarray}    
where $k$ is the quasimomentum, $j$ is the band index and 
the function $\tilde{\varphi}_{jk}(z)$ is periodic with period $d$. 
Notice that Eq.(\ref{bloch-ansatz}) does not exhaust all the possible stationary solutions of the Gross-Pitaevskii equation \cite{machholm2}.      
The Gross-Pitaevskii equation (\ref{gpe}), rewritten in terms of   
the functions $\tilde{\varphi}_{jk}(z)$, reads    
  
\begin{eqnarray}    
\Bigg[ {1 \over 2m}     
\left( -i \hbar\partial_z + k \right)^2    
+s\;E_R\;{\rm sin}^2\left({\pi z \over d }\right) +      
\nonumber \\     
 + gn d  |\tilde{\varphi}_{jk}(z)|^2  \Bigg]  \tilde{\varphi}_{jk}(z)     
= \mu_j(k)  \tilde{\varphi}_{jk}(z).  
\label{galileo}    
\end{eqnarray}    
From the solution of Eq.(\ref{galileo}) one gets the functions $\tilde{\varphi}_{jk}(z)$ and the corresponding chemical potentials $\mu_j(k)$.

The energy per particle $\varepsilon_j(k)$ can be calculated using the expression
\begin{eqnarray}    
{\varepsilon_j } (k)= \int_{-d/2}^{d/2}     
\tilde{\varphi}^*_{jk}(z) \left[ {1 \over 2m }    
\left(  -i \hbar\partial_z + k \right)^2    
+ \right. \nonumber\\    
\left.     
+ s\;E_R \; {\rm sin}^2\left({z}\right)     
+ {1 \over 2}{gn d } |\tilde{\varphi}_{jk}(z)|^2 \right]  \tilde{\varphi}_{jk}(z) dz.    
\label{epsilon_l}    
\end{eqnarray}    
and differs from the chemical potential $\mu_j(k)$. 
In fact by multiplying Eq.(\ref{galileo}) by $\tilde{\varphi}_{jk}^*$ and integrating, one finds the expression
 \begin{eqnarray}    
{\mu_j } (k)= \int_{-d/2}^{d/2}     
\tilde{\varphi}^*_{jk}(z) \left[ {1 \over 2m }    
\left(  -i \hbar\partial_z + k \right)^2    
+ \right. \nonumber\\    
\left.     
+ s\;E_R \; {\rm sin}^2\left({z}\right)     
+ {gn d } |\tilde{\varphi}_{jk}(z)|^2 \right]  \tilde{\varphi}_{jk}(z) dz.    
\label{mubands}
\end{eqnarray} 
for the chemical potential 
which coincides with Eq.(\ref{epsilon_l}) only in absence of the interaction term. 
In general, $\mu_j$ and $\varepsilon_j$ are related to each other
by Eq.(\ref{mu_l}).

The solution of Eq.(\ref{galileo}) for $k=0$ and $j=1$ gives the
ground state of the system.  This state corresponds to a condensate at
rest in the frame of the optical lattice.  Instead the solutions of
Eq.(\ref{galileo}) with $k\neq 0$, describe states of the system where
all the atoms, occupying the same single-particle wavefunction,
move together with respect to the optical potential giving rise to the
constant current (\ref{current}).
Experimentally such states can be created by turning on adiabatically the intensity
of a lattice moving at fixed velocity \cite{denschlag}.  In this way, 
it is possible to map higher Brillouin zones onto higher bands.  
  
\begin{center}    
\begin{figure}    
\includegraphics[width=0.8\linewidth]{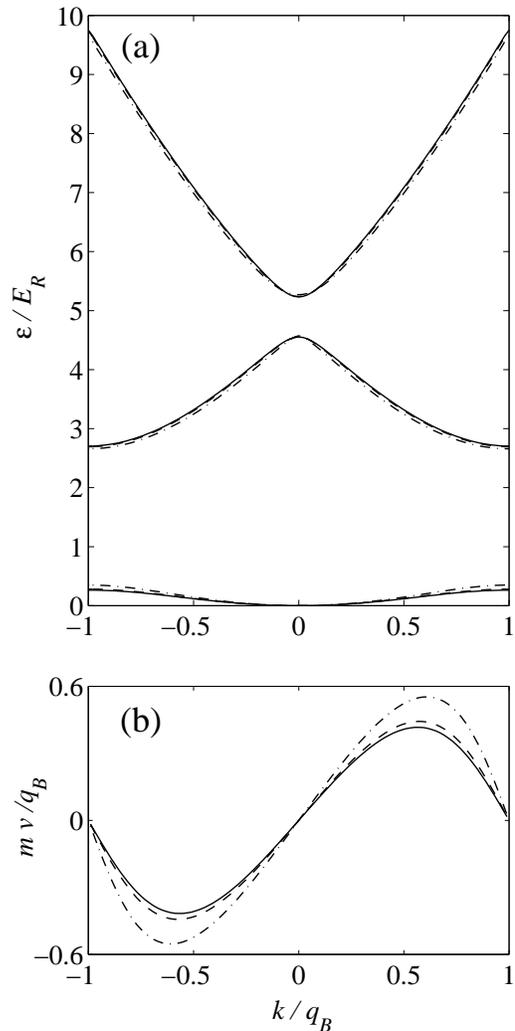}    
\caption{(a) Lowest three Bloch bands for $s=5$, $gn=0$ (solid line),
$0.1 E_R$ (dashed line) and $0.5 E_R$ (dash-dotted line). The energy of the groundstate ($k=0$) has been subtracted; (b) Group
velocity for the same parameters as in (a).}
\label{bloch-bands}
\end{figure}    
\end{center}

Results for the Bloch bands (\ref{epsilon_l}) are shown in Fig.\ref{bloch-bands} for
$s=5$ and $gn/E_R=0,\,0.1$ and $0.5$. The effect of
interactions for these parameter values can be hardly distinguished in the
energy band structure.  It can be made more evident by plotting
the group velocity
 
\begin{eqnarray}
v={\partial \varepsilon(k) \over \partial k}
\label{groupv}
\end{eqnarray}
as a function of the quasi-momentum. For $gn=0.5 E_R$ one finds a
difference in the group velocity of about 30\% with respect to the non
interacting case.  The quantity plotted in
Fig.\ref{bloch-bands}(b) is accessible experimentally through Bloch
oscillations experiments \cite{arimondo}.
In this context, it is important to note that the Bloch states (\ref{bloch-ansatz}) with $k\neq 0$ can become energetically or dynamically unstable depending on the choice of the values for $s$ and $gn$ (See for example \cite{niu,smerzi,machholm,trombettoni,menotti}).
Note also that for values of $gn \ge s$ 
one encounters loops (``swallow tails'') in the band structure at the band edge of the lowest band 
\cite{machholm,swallow} and, at even smaller values of $gn$, 
at the center of the first excited band \cite{machholm}. 
For the values of $gn$ we have considered, these swallow tails exist only for 
very small values of the lattice depth $s$.

\begin{center}    
\begin{figure}    
\includegraphics[width=0.8\linewidth]{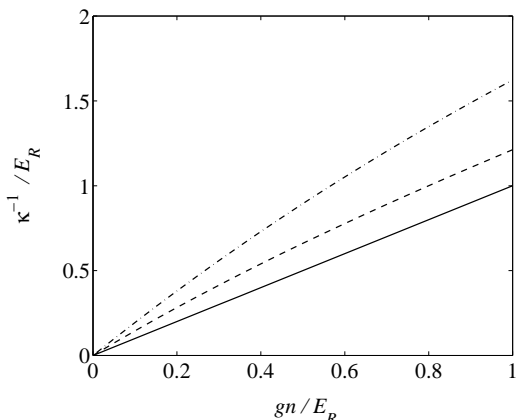}    
\caption{Inverse compressibility $\kappa^{-1 } = n {\partial \mu /
\partial n}$ as a function of $g n/E_R$ for $s=0$ (solid line), $s=5$
(dashed line) and $s=10$ (dashed-dotted line).}
\label{fig_inv_compr}    
\end{figure}    
\end{center}

\begin{center}    
\begin{figure}    
\includegraphics[width=0.8\linewidth]{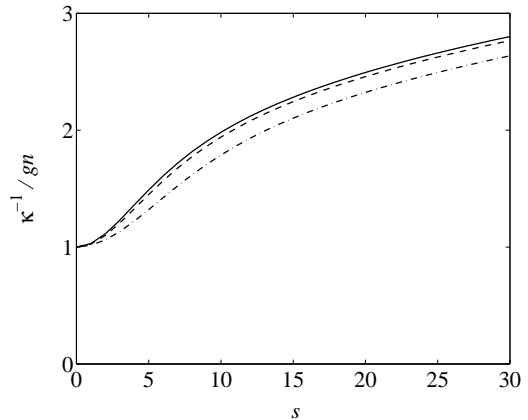}    
\caption{$\kappa^{-1}/gn$ for $gn=0.1 E_R$ (dashed line) and $gn=0.5
E_R$ (dashed-dotted line) as a function of the lattice depth $s$;  
comparison with the effective coupling constant $\tilde{g}/g$ defined in
Eq.(\ref{gtilde}) (solid line).}
\label{fig_gtilde}    
\end{figure}    
\end{center}

Let us now focus on the properties of the ground state ($k=0,\,j=1$).  By
solving numerically Eq.(\ref{galileo}) and
calculating the chemical potential, it is straightforward to evaluate
the inverse compressibility (\ref{kappa}) as a function
of the relevant parameters of the problem.  The results are plotted in
Fig.\ref{fig_inv_compr} as a function of $gn$ for $s=0,5,10$.  The
case $s=0$ is the uniform case, where the equation
of state is $\mu=gn$ and $\kappa^{-1 }=g n$.  In the presence of the
optical lattice, we predict a deviation from this linear dependence on density. One
finds an increase of the inverse compressibility with $s$, which is a
direct consequence of the localization of the wavefunction at the
bottom of the wells produced by the optical lattice.  When the effect
of two body interactions on the wavefunction is negligible one can
account for this increase through the simple law $\kappa^{-1}={\tilde
g}(s) n$.  The effective coupling constant ${\tilde g}(s)$ depends
only on the lattice depth $s$ and takes the explicit form 

\begin{eqnarray}
{\tilde g}=
g d \int_{-d/2}^{d/2} \varphi_{gn=0}^4 dz ,
\label{gtilde}
\end{eqnarray}
where $\varphi_{gn=0}$ is the groundstate solution of Eq.(\ref{gpe})
for $gn=0$.
It is correct to describe $\kappa^{-1}$ with a linear dependence on
the density only for small interaction parameters $gn$.  For higher
values of the interactions the slope of the curves tends to
decrease. This is due to the fact that interactions tend to
broaden the order parameter in each well and hence counteract the effect 
produced by the optical lattice.
Numerical results for $\kappa^{-1}/gn$ are presented in
Fig.(\ref{fig_gtilde}) and compared with the density independent quantity
${\tilde g}/g$ (solid line), confirming that in general the compressibility does not 
depend linearly on the interaction. However for large $s$ the
density dependence of $\kappa^{-1}/gn$ becomes less and less
important and the expression $\kappa^{-1}(n,s)=\tilde{g}(s)n$ becomes
applicable in a larger range of $gn$-values.

\begin{center}    
\begin{figure}    
\includegraphics[width=0.8\linewidth]{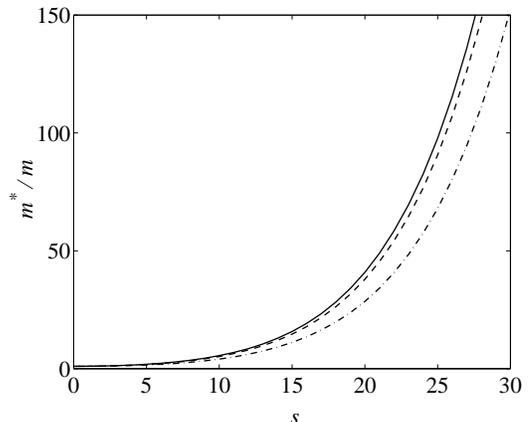}    
\caption{Effective mass as a function of lattice depth $s$ for $gn=0$
(solid line), $gn=0.1 E_R$ (dashed line) and $gn=0.5 E_R$
(dashed-dotted line).}
\label{fig_eff_mass}    
\end{figure}    
\end{center}

Let us now determine the effective mass by studying the low-$k$
behaviour of the lowest band $\varepsilon(k)$.  The results for
$m^*=m^*(k=0)$ are shown in Fig.\ref{fig_eff_mass}.  For $s
\to 0$, the effective mass tends to the bare mass $m$.  Instead for
large $s$ the effective mass increases strongly due to the
decreased tunneling between neighbouring wells of the optical
potential. The effect of the interactions is to decrease the value of
$m^*$ as a consequence of the broadening of the wavefunction caused by
the repulsion, which favours tunneling, contrasting the effect of the lattice potential.  
In fact, the effective mass is fixed by the
tunneling properties of the system, which are exponentially sensible
to the behaviour of the wavefunction in the region of the barriers.  Then,
any small change in the wavefunction due to interactions can have a
significant effect on the effective mass.

\begin{center}    
\begin{figure}    
\includegraphics[width=0.8\linewidth]{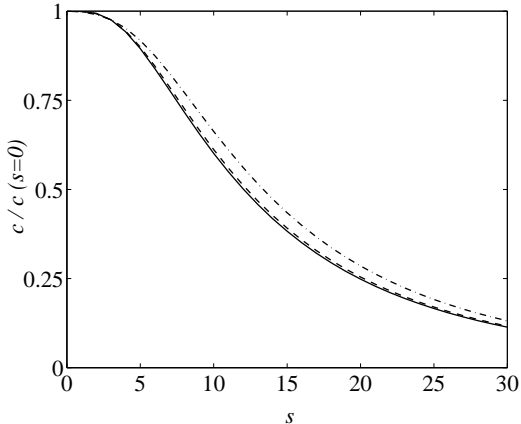}    
\caption{Sound velocity as a function of the potential depth $s$
divided by the sound velocity in the absence of the optical potential
($s=0$) for $gn=0.02 E_R$ (solid line), $gn=0.1 E_R$ (dashed line) and
$gn=0.5 E_R$ (dashed-dotted line).}
\label{fig_sound_vel}    
\end{figure}    
\end{center}

The two quantities calculated above, compressibility and effective
mass, can be used to calculate the sound velocity using relation
(\ref{sound_vel}) \cite{machholm,meret}.  The corresponding results are shown in
Fig.\ref{fig_sound_vel}.  We find that the sound velocity decreases as
the lattice is made deeper.  This is due to the fact that the increase
of the effective mass is more important than the decrease of the
compressibility $\kappa$.  The solid line in
Fig.\ref{fig_sound_vel} shows that decreasing the interactions the
sound velocity approaches the law $c=\sqrt{\tilde{g}n/m^*_{gn=0}}$, where $m^*_{gn=0}$ is the effective mass calculated with $gn=0$.
The sound velocity is in principle measurable by studying the
velocity of a wavepacket propagating in the presence of the optical
potential. This could be done for example by following the experimental 
procedure used in \cite{andrews}. 
Yet note that in deep lattices, non linear effects are expected to 
be important also for small amplitude perturbations \cite{smerzi}.
      
Most of the results discussed above can be qualitatively understood by
working in the so called tight binding approximation which becomes
more and more accurate as the intensity of the laser field increases.
Within the tight binding approximation we can derive analytic or
semi-analytic expressions for the energy bands and consequently for
the effective mass, compressibility and sound velocity.
  
Using Bloch's theorem, we can write the condensate in the lowest band   
as

\begin{eqnarray}  
\varphi_k(z)= \sum_l e^{i k l d/\hbar } f(z-l d),  
\label{t-b-varphi}  
\end{eqnarray}  
where $l$ is the index of the well and $f$ are the Wannier functions.
The function $f(z)$ is normalised to unity and is orthogonal to the functions
centered at different sites. 
In general, it depends on two-body interactions and hence on the density. 
Equation (\ref{t-b-varphi}) holds for any depth of the optical potential.
However, in the tight binding regime, $f(z)$ is a well localized
function. This provides important simplifications in the calculation of
the relevant quantities, since only nearest-neighbour overlap integrals
have to be considered.
  
In our derivation of the tight binding results we will include also
the interaction terms in the Hamiltonian. By substituting
Eq.(\ref{t-b-varphi}) into (\ref{galileo}) and using definition
(\ref{epsilon_l}) it follows, after some straightforward algebra, that
the energy per particle takes the simple tight-binding form

\begin{eqnarray}    
\varepsilon(k) =
\varepsilon-      
\delta \;{\rm cos}  \left({k d \over \hbar}\right),  
\label{t-b}    
\end{eqnarray}    
where $\varepsilon=\int f(z) \left[ -{\hbar^2 \partial^2_z \over 2m}
+V(z)+ {gnd\over 2} f^2(z) \right] f(z) dz$ is an energy offset, which
depends on $s$ and $gn$ but not on $k$, and $\delta$ is the tunneling
parameter defined as
  
\begin{eqnarray}  
\delta =-2\! \! \int\!\! f(z)\!\!    
\left[ -{\hbar^2  \partial^2_z    \over 2m}  
+V(z)+ 2 gnd f^2(z) \right]\! \! f(z\!-\!d) dz.  
\label{delta_e}  
\end{eqnarray}  
Using the same approximations, the chemical potential takes the form  
  
\begin{eqnarray}  
\mu(k) = \mu_0- \delta_\mu \cos \left({kd\over\hbar}\right), 
\label{t-b-mu}  
\end{eqnarray}  
where $\mu_0= \int f(z) \left[ -{\hbar^2  \partial^2_z    \over 2m}  
+V(z)+  gnd f^2(z) \right] f(z) dz$ 
and $\delta_{\mu} = \delta - 4 gnd \int f^3(z)  f(z-d) dz$.  
To derive Eqs.(\ref{t-b},\ref{t-b-mu}), we have kept terms of the
order $\int f^3(z)f(z-d) dz$ and neglected terms of the order $\int
f^2(z)f^2(z-d) dz$, which, for localized functions, turn out to be
much smaller.

Note that $\delta$ and $\delta_{\mu}$ depend on density both explicitly and 
implicitly through the density dependence of $f$ (see Eq.(\ref{delta_e})). 
In all situations considered in this work, contributions involving $\partial f/\partial n$ can be safely neglected. 
This approximation allows us to identify the quantity $\delta_\mu-\delta = -4 gnd \int f^3(z)f(x-d)$ with $n\partial\delta/\partial n$.

\begin{center}    
\begin{figure}    
\includegraphics[width=0.8\linewidth]{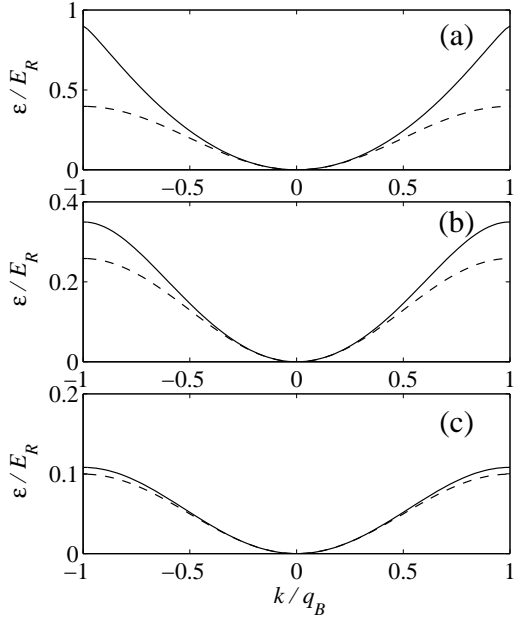}    
\caption{Lowest Bloch band at $gn=0.5 E_R$ for different values of the
potential depths: $s=1$ (a), $s=5$ (b) and $s=10$ (c).  The solid
lines are obtained by evaluating Eq.(\ref{epsilon_l}) using the
numerical solution of Eq.(\ref{galileo}) while the dashed lines refer
to the tight binding expression (\ref{t-b}). The energy of the groundstate 
($k=0$) has been subtracted.}
\label{t-b_gn05}    
\end{figure}    
\end{center}

To check the accuracy of the tight binding approximation, one can
compare the numerical results for the first Bloch band with expression
(\ref{t-b}). The parameter $\delta$ is related to the curvature of the
band at $k=0$, i.e. to the effective mass defined in
Eq.(\ref{curvature}), through the important relation \cite{notedeltamu} 

\begin{eqnarray}    
\delta={2 \over \pi^2}{m \over m^*} E_R .    
\label{delta}    
\end{eqnarray}    
Hence, we can evaluate (\ref{t-b}) by using the numerical results for the 
effective mass discussed above (see Fig.(\ref{fig_eff_mass})). In this way, we automatically include the correct density dependence of the tunneling parameter $\delta$. 
In Fig.\ref{t-b_gn05}, we compare the first Bloch energy band with its
tight binding approximation for $gn=0.5 E_R$ and various values of
$s$. The comparison shows that, for this value of the interactions,
the tight binding approximation is already quite good for $s=10$ and
becomes better and better for increasing $s$.
  
Using Eq.(\ref{t-b-mu}) for $k=0$, we can also derive an expression for   
the  inverse compressibility, which reads  
  
\begin{eqnarray}  
\kappa^{-1}= gnd \int f^4(z) dz + 8gnd \int f^3(z)f(z-d) dz,  
\label{t-b-k}  
\end{eqnarray}  
where contributions due to $\partial f/\partial n$ have been neglected, as previously.  
Since $\int f^3(z)f(z-d) dz$ is normally much smaller than $ \int
f^4(z) dz$, the on-site contribution to the inverse compressibility
(first term in (\ref{t-b-k})) is usually the leading term.

A tight-binding expression for the density-independent 
effective coupling constant $\tilde{g}$ (\ref{gtilde}) 
is obtained from (\ref{t-b-k}) by 
replacing $f$ with the Wannier function $f_{gn=0}$ of the non-interacting system. 
Then $\kappa^{-1}$ is linear in the density and $\kappa^{-1}/n$ can be identified with $\tilde{g}$. 
Neglecting the overlap contribution, $\tilde{g}$ takes the form \cite{meret}

\begin{eqnarray}
{\tilde g}=g d \int f^4_{gn=0}(z)dz \,.
\label{gtilde1}
\end{eqnarray}
This shows that in the tight binding regime, the effective coupling constant $\tilde{g}$ can be safely estimated replacing the Bloch state $\varphi_{gn=0}$ in Eq.(\ref{gtilde}) by the Wannier function $f_{gn=0}$.

Deep in the tight binding regime, the function $f(z)$ can be
conveniently approximated by a gaussian
$f=\exp(-z^2/2\sigma^2)/\pi^{1/4}\sqrt{\sigma}$, where the width $\sigma$
is chosen such as to satisfy the equation

\begin{equation}
-{d^3\over\pi^3}{1\over\sigma^3}+s{\pi\over d}\sigma-s{\pi^3\over d^3}\sigma^3
-{1\over 2}{gn\over E_R}\sqrt{\pi\over 2}{d^2\over \pi^2}{1\over\sigma^2}
=0
\label{def-sigma}
\end{equation}
accounting for the anharmonicities of ${\cal{O}}(z^4)$ of the
potential wells and for the broadening effect of repulsive two body
interactions.  Within the gaussian approximation, the inverse
compressibility (\ref{t-b-k}) can be rewritten in the simplified form

\begin{eqnarray}  
\kappa^{-1}= {gnd\over\sqrt{2\pi}\sigma}\,,
\label{t-b-k-g}  
\end{eqnarray}  
where we have neglected the contribution arising from the overlap of
neighbouring wavefunctions.  If we neglect the interaction term in
Eq. (\ref{def-sigma}), we obtain $\kappa^{-1}= \tilde{g}n$, with
$\tilde{g}={gd/\sqrt{2\pi}\sigma}$ and $\sigma=s^{-1/4}(1+1/4\sqrt{s})d/\pi$. 
For $s=10, gn=0.5 E_R$ the approximation (\ref{t-b-k-g}) differs from 
the exact value of $\kappa^{-1}$ by less than $1\%$.

Note that even though the gaussian approximation is useful in
estimating the compressibility at high $s$, it can not be employed to
calculate the effective mass or, equivalently, the tunneling parameter
$\delta$, which requires a more accurate description of the tails 
of the function $f$.

\section{Elementary excitations}    
\label{elementary-excitations}

In this section we study the spectrum of elementary excitations.  
We will focus our attention on the excitations  
relative to the ground state, i.e. to the solution  
of Eq.(\ref{galileo}) with $j=1$ (lowest band) and $k=0$.  
To this aim we look for solutions 
of the time-dependent Gross-Pitaevskii equation  
of the form  
    
\begin{eqnarray}
\varphi(z,t)=e^{-i \mu t / \hbar}
\left[
\varphi(z) + u_{jq}(z) e^{-i\omega_j(q)t} + v^*_{jq}(z) e^{i\omega_j(q)t}
\right],
\end{eqnarray}
where $u_{jq}$ and $v_{jq}$ describe a small perturbation with respect to the groundstate 
condensate $\varphi\equiv \varphi_{j=1,k=0}$. 
At first order in the perturbations, 
the time-dependent Gross-Pitaevskii equation 
yields the Bogoliubov equations

\begin{eqnarray}    
\label{bog_u}    
\left[      
-{\hbar^2 \partial^2_z   \over 2m}   
+ s \; E_R \; {\rm sin}^2\left({\pi z \over d }\right)    
- \mu + 2 gn d |\varphi|^2 \right] u_{jq}(z) + \nonumber \\    
+ gnd  \varphi^2  v_{jq}(z) = \hbar \omega_j(q) u_{jq}(z), \\    
\left[    
-{\hbar^2 \partial^2_z   \over 2m}   
+ s \;E_R \;  {\rm sin}^2\left({\pi z \over d }\right)    
- \mu + 2 gnd |\varphi|^2 \right] v_{jq}(z) + \nonumber \\    
+ gn d \varphi^{*2}  u_{jq}(z) = - \hbar \omega_j(q) v_{jq}(z).
\label{bog_v}    
\end{eqnarray}    
The solutions $u_{jq}$ and $v_{jq}$ are Bloch waves ($u_{jq}=\exp(iqz/\hbar)
\tilde{u}_{jq}(z)$ where ${\tilde u}_{jq}$ is periodic with period $d$
and analogously for $v_{jq}$). 
They are labeled by their band
index $j$ and their quasimomentum $q$ belonging to the first
Brillouin zone. 
Hence, also the Bogoliubov spectrum $\omega_j(q)$ 
exhibits a band structure \cite{moelmer,chiofalo}.
   
\begin{center}    
\begin{figure}    
\includegraphics[width=0.8\linewidth]{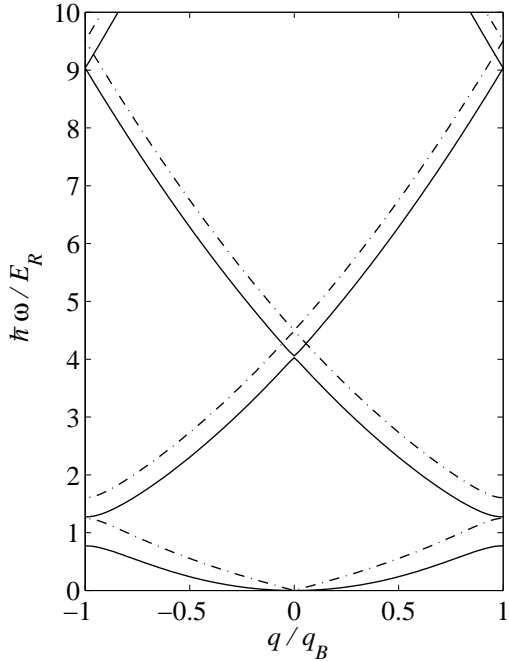}
\caption{Bogoliubov bands in the first Brillouin zone for $s=1$,
$gn=0$ (solid line) and $gn=0.5 E_R$
(dash-dotted line). Note that for such a small potential, the gap
between second and third band is still very small. }
\label{effect-interactions-s5}    
\end{figure}    
\end{center}    

Similarities and differences with respect to the well-known Bogoliubov
spectrum in the uniform case ($s=0$) are immediate.  As in the uniform
case, interactions make the compressibility finite, giving rise to a
phononic regime for long wavelength excitations ($q\to 0$) in the
lowest band.  In high bands the spectrum of excitations instead resembles the
Bloch dispersion (see Eq.(\ref{epsilon_l})). 
The differences are due to
the fact that in the presence of the optical lattice 
the Bogoliubov spectrum develops a band structure.  As a
consequence, the dispersion is periodic as a function of quasimomentum
and different bands are separated by an energy gap.  In particular,
the phononic regime present at $q=0$ is repeated at every even
multiple of the Bragg momentum $q_B$. Moreover, the lattice period $d$
emerges as an additional physical length scale.

In Fig.\ref{effect-interactions-s5} we compare the Bologoliubov bands at 
$s=1$ for $gn=0$ and $gn=0.5 E_R$.  In the interacting case, 
one notices the appearance of the phononic regime in the lowest band, 
while higher bands differ from the non-interacting ones mainly by an energy shift.

\begin{center}    
\begin{figure}    
\includegraphics[width=0.8\linewidth]{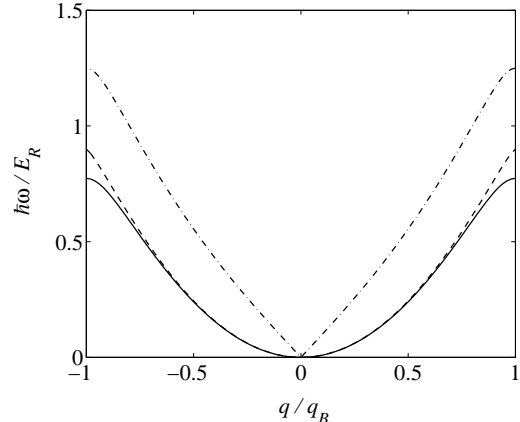}    
\caption{Lowest Bloch band (dashed line) and lowest Bogoliubov band
(dash-dotted line) bands for $s=1$ and $gn=0.5 E_R$ compared with
the energy band without interactions for $s=1$ (solid line). 
The groundstate energy has been subtracted in the case of the Bloch band 
(dashed line) and of the energy band without interactions (solid line).}
\label{fig-bog-bloch}    
\end{figure}    
\end{center}

In Fig.\ref{fig-bog-bloch} we compare the lowest Bogoliubov and Bloch
bands. Clearly, the lowest Bloch band is less affected by the presence
of interactions than the Bogoliubov band.  Recall that the Bogoliubov
band gives the energy of the elementary excitations while the
Bloch band gives the energy per particle of an excitation involving
the whole condensate.

\begin{center}    
\begin{figure}    
\includegraphics[width=0.8\linewidth]{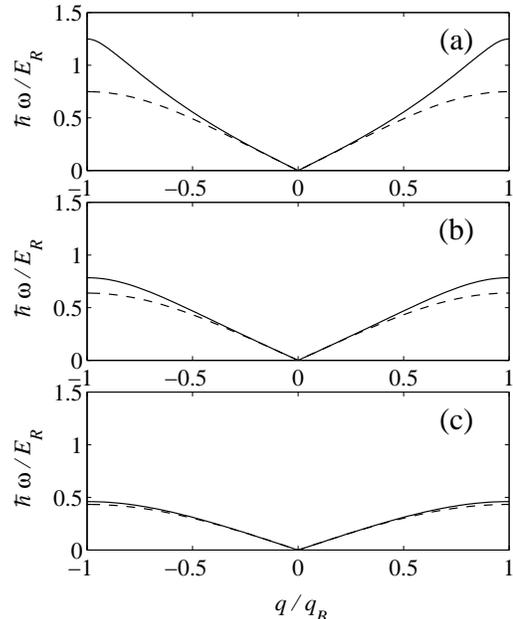}    
\caption{Lowest Bogoliubov band at $gn=0.5 E_R$ for different values
of the potential depths: $s=1$ (a), $s=5$ (b) and $s=10$ (c).  
The solid lines are obtained from the numerical solution of
Eqs.(\ref{bog_u},\ref{bog_v}) while the dashed lines refer to the
tight-binding expression (\ref{approx_bog_t-b}). }
\label{fig-bog-tba-2}    
\end{figure}    
\end{center}

The solid lines in Fig.\ref{fig-bog-tba-2} show how the lowest
Bogoliubov band changes when the lattice depth is increased at fixed
interaction.  At $s=1$ (Fig.\ref{fig-bog-tba-2}a), apart from the formation of the energy gap
close to $q=q_B$, the curve still resembles the dispersion in the
uniform case: both the phononic linear regime and the quadratic regime
are visible.  When the potential is made deeper ($s=5, 10$; Fig.\ref{fig-bog-tba-2}b,c), 
the band
becomes flatter.  As a consequence, the quadratic regime disappears
and the slope of the phononic regime decreases. This reflects the
behaviour of the velocity of sound (see Fig.\ref{fig_sound_vel}). 

Similarly to the Bloch energy and chemical potential spectra, also for
the Bogoliubov excitation spectrum we can obtain an analytic expression
in the tight binding limit.  We write the excitation amplitudes in the
form

\begin{eqnarray}  
\label{t-b-u}  
u_q(z)= U_q \sum_l e^{i q l d / \hbar} f(z-ld), \\  
v_q(z)= V_q \sum_l e^{i q l d / \hbar} f(z-ld),  
\label{t-b-v}  
\end{eqnarray}  
where $f$ is the same function as in (\ref{t-b-varphi}). Recall, that $\int dz f^2=1$.
Using expression (\ref{t-b-varphi}) for the order parameter, we
find the result

\begin{eqnarray}    
\label{approx_bog_t-b}    
\hbar \omega_q \!=\! \hspace*{6cm} \\ 
\sqrt{ 2\delta{\rm sin}^2\left({q d \over 2 \hbar }\right) 
\left[ 2 \left(\delta + 2n{\partial\delta\over\partial n}\right)
{\rm sin}^2 \left({q d \over 2 \hbar}\right) + 2 \kappa^{-1}\right]}
\nonumber
\end{eqnarray}    
for the excitation frequencies (lowest band), where $\delta$ and
$\kappa^{-1}$ are the tunneling and inverse compressibility parameters
defined respectively in (\ref{delta_e}) and (\ref{t-b-k}).
In the same limit, the Bogoliubov amplitudes are 

\begin{eqnarray}
U_q &=&
\frac{ \varepsilon_q + \hbar \omega_q}
{2 \sqrt{ \hbar \omega_q \varepsilon_q}},  \label{U_tb}
\\
V_{q}&=& 
\frac{ \varepsilon_q - \hbar \omega_q}
{2 \sqrt{ \hbar \omega_q \varepsilon_q}},
\label{V_tb}
\end{eqnarray}
where $\omega_q$ is given by Eq.(\ref{approx_bog_t-b}) and
$\varepsilon_q= 2\delta{\rm sin}^2\left({q d / 2 \hbar }\right)$
captures the quasimomentum dependence of the Bloch energy (compare with
Eq.(\ref{t-b})). 
Note that in deriving Eqs.(\ref{U_tb},\ref{V_tb}), we have imposed the normalization 
condition $\int_{-d/2}^{d/2} |u_{q}(z)|^2-|v_{q}(z)|^2 dz =1$.

In expression (\ref{approx_bog_t-b}) the density dependence 
of the spectrum shows up in three different ways:   
  
- first of all, the parameter $\delta$ depends on interactions
  as shown explicitly in Fig.\ref{fig_eff_mass}, where the quantity
  $m^*\propto 1/\delta$ (see relation (\ref{delta})) is plotted;

- second, $\kappa^{-1}$ has a more general dependence on the density
  than the one accounted for by the linear law $\tilde{g}n$, as shown
  in Fig.\ref{fig_gtilde};

- third, a contribution due to the density derivative of $\delta$
appears. However this term is always small: for small
interactions one has $n \partial\delta/\partial n \ll \delta$ while for larger
interactions the inverse compressibility $\kappa^{-1}$ dominates both
$\delta$ and $n \partial\delta/\partial n$.  However, as shown in
\cite{trombettoni,menotti} this term can significantly affect the
excitation frequency calculated on top of a moving condensate.

Fig.\ref{fig-bog-tba-2} compares the numerical data with the approximate expression 
(\ref{approx_bog_t-b}) evaluated using the parameters $\kappa^{-1}$
and $\delta$ calculated in the previous section. The tunneling
parameter $\delta$ is obtained from the data for the effective mass
$m^*$ through Eq.(\ref{delta}). As already found for the lowest Bloch
band, for this value of $gn$, the agreement with the tight
binding expression is already good for $s=10$.

It is possible to identify two regimes, where the
Bogoliubov spectrum can be described by further simplified expressions:

(I) for very large potential depth, the spectrum is dominated by the second term
in the square brackets of Eq.(\ref{approx_bog_t-b}). 
In fact, $\delta\rightarrow 0$ while $\kappa^{-1}$ becomes larger and larger as $s$ increases. 
Hence, the spectrum takes the form

\begin{eqnarray}    
\hbar \omega_q \approx  
\sqrt{ \delta \kappa^{-1}} 
\left|  \sin\left({q d \over 2 \hbar }\right) \right|\,,
\label{bog-tb-larges}
\end{eqnarray}
both for large and small $gn$.  Of course for large $gn$, the proper
density-dependence of $\delta$ and $\kappa^{-1}$ has to be taken into account in evaluating (\ref{bog-tb-larges}).  
Note that the Bogoliubov band becomes very flat since $\delta$ decreases exponentially for large lattice depth $s$. 
Yet, its height decreases more slowly than the lowest Bloch band (\ref{t-b}) whose height decreases linearly in $\delta$.
We also
point out that in the regime where Eq.(\ref{bog-tb-larges}) is valid, the Bogoliubov amplitudes Eqs.(\ref{U_tb},\ref{V_tb}) become
comparable in magnitude for all $q$ in the first Brillouin zone, even if the
excitation spectrum(\ref{bog-tb-larges}) is not
linear. This implies that all excitations in the lowest band acquire
quasi-particle character and the role of interactions is strongly enhanced.

(II) for small enough $gn$, one can neglect the density dependence of
$\delta$ and use the approximation $\kappa^{-1}={\tilde g} n$, where
${\tilde g}$ was defined in (\ref{gtilde}) and takes the form (\ref{gtilde1}) 
in the tight binding regime. This yields
\begin{eqnarray}    
\hbar \omega_q \approx    
\sqrt{ 2 \; \delta_0  \; {\rm sin}^2  \left({q d \over 2 \hbar }\right)    
\left[ 2 \; \delta_0\; {\rm sin}^2  \left({q d \over 2 \hbar }\right)    
 + 2 \tilde{g}n\right]}\,,
\label{i}  
\end{eqnarray}    
which was first obtained in \cite{javanainen} (see also \cite{smerzi,oosten,rey}). 
Eq.(\ref{i}) has a form similar to the well-known Bogoliubov spectrum of uniform gases, the energy $2\delta_0 {\rm sin}^2  \left({q d /2 \hbar }\right)$ replacing the free particle energy $q^2/2m$.
    
The spectrum of elementary excitations can be measured by exposing the
system to a weak perturbation transferring momentum $p$ and energy
$\hbar\omega$.  The response of the system is described by the dynamic
structure factor $S(p,\omega)$ which, in the presence of a periodic
potential takes the form

\begin{eqnarray}
S(p,\omega) = \sum_j Z_j(p) \delta(\omega - \omega_{j}(p)),
\label{struc-fac}
\end{eqnarray} 
where $Z_j(p)$ are the density excitation strengths relative to the $j^{th}$
band and $\hbar \omega_{j}(p)$ are the
corresponding excitation energies, defined by the solutions of
Eqs.(\ref{bog_u},\ref{bog_v}).  
Note that $p$, here assumed to be
along the optical lattice ($z$ axis), is not restricted to the first
Brillouin zone, being the physical momentum transferred to the system by the external
probe. In this respect, it is important to notice that, while the
excitation energies $\hbar \omega_{j}(p)$ are periodic as a function of
$p$, this is not true for the strengths $Z_j(p)$.  
Starting
from the solution of Eqs.(\ref{bog_u}) and (\ref{bog_v}), the 
excitation strengths $Z_j(p)$ 
\pagebreak
can be evaluated using the 
standard prescription of Bogoliubov theory (see
for example \cite{griffin})

\begin{eqnarray}
Z_j(p) = \left| \int_{-d/2}^{d/2} 
\left[ u^*_{jq}(z) + v^*_{jq}(z) \right] e^{ipz/\hbar} \varphi(z) 
dz \right|^2 ,
\label{strength}
\end{eqnarray}
where $q$ belongs to the first Brillouin zone and is fixed by the
relation $q=p+2 \ell q_B$ with $\ell$ integer and the Bogoliubov amplitudes are normalized 
according to $\int_{-d/2}^{d/2} |u_{jq}(z)|^2-|v_{jq}(z)|^2 dz =1$.  The dynamic structure
factor of a Bogoliubov gas in an optical lattice 
has been recently calculated in \cite{chiara}.  It is found
that for an interacting system the strength towards the
first band develops an oscillating behaviour as a function of the
momentum transfer, vanishing at even multiples of the Bragg momentum
due to the presence of a phononic regime. Moreover, in the presence of
interactions, the strength $Z_1$ towards the first bands vanishes as
$\sqrt{\delta \kappa}$ in the limit of very deep lattices, due to the
quasi-particle character of the excitations in the whole band.
Finally, the suppression of the static structure factor at small
momenta, due to phononic correlations, is significantly
enhanced by the presence of the lattice.

\subsection{Link with the Josephson formalism}  
\label{josephson}

In this section, we show that the results for the excitation spectrum in the tight binding limit 
obtained in the previous section (see Eq.(\ref{approx_bog_t-b})) can
be recovered using a different formalism based on the Josephson
equations of motion.  These can be derived starting from the ansatz
for the condensate 

\begin{eqnarray}    
\varphi(z,t)=\sum_l\,f_l(z;n_l)\,\sqrt{n_l(t)\over n}\,e^{iS_l(t)},    
\label{tb_ansatz}    
\end{eqnarray}    
where for sufficiently deep optical lattices the wavefunction $f_l$ is
localized at  site $l$ and extends only over nearest-neighbouring
sites, $n_l$ is the time-dependent average density at site $l$ 
where the average is taken over the $l$-th well, $n$ is the average equilibrium density and 
 $S_l$ is the phase of the condensate at site $l$.  At equilibrium the functions
$f_l$ coincide with the Wannier functions $f(z-ld)$ introduced before. In
general, they depend on the density at the corresponding site, as
indicated in Eq. (\ref{tb_ansatz}), and hence might themselves
implicitly depend on time.  Furthermore, these functions are chosen
such that $\int f_{l'}^*(z;n_{l'}) f_{l}(z;n_{l})dz=0$ for $l\neq l'$ and 
$1$ for $l=l'$.
  
When excitations are present, the phases $S_l$ at different sites will
be different from each other, indicating the presence of a current.
Using the time-dependent Gross-Pitaevskii formalism one can
derive the following equations of motion for the density and phase
variables

\begin{eqnarray}  
{\dot n_l}&=&\sum_{l'=l+1,l-1}\;{\delta^{l,l'}\over\hbar}
\sqrt{n_l n_{l'}}\;\sin(S_l-S_{l'})\,,
\label{jos1} \\  
{\dot S_l}&=&-{\mu_l\over \hbar}
+\sum_{l'=l+1,l-1}\;{\delta_{\mu}^{l,l'}\over 2\hbar}
\sqrt{n_{l'} \over n_l}\;\cos(S_l-S_{l'})\,,
\label{jos2}  
\end{eqnarray}    
where $\mu_l= \int f_l \left[ -{\hbar^2 \partial^2_z \over 2m} +V(z)+
gnd |f_l|^2 \right] f_l dz$, while the time-dependent tunneling
parameters $\delta^{l,l'}$ and $\delta_{\mu}^{l,l'}$ are directly
related to the overlap between two neighbouring wavefunctions

\begin{eqnarray}    
\label{delta-fm}    
\delta^{l,l'}&=&   
-2\!\!\int \!dz    
\Bigg[    
f_l \left( - {\hbar^2 \partial_z^2\over 2m} \!+    
  V  \right)f_{l'}  + \\    
&&   
+gn_ld\,f_l|f_l|^2f_{l'}  
+gn_{l'}d\,f_l|f_{l'}|^2f_{l'}  
\Bigg]\,,  \nonumber  \\  
\delta_{\mu}^{l,l'} &=& \delta^{l,l'}   
-4 gn_l d  \int f_l|f_l|^2f_{l'} dz.  
\label{delta_mu-fm}    
\end{eqnarray}    
Note that at equilibrium $\delta^{l,l'}=\delta$ and
$\delta_{\mu}^{l,l'}=\delta_{\mu}$, where $\delta$ and $\delta_{\mu}$
have been previously defined in Eq.(\ref{delta_e}) and after
Eq.(\ref{t-b-mu}).  At equilibrium they do not depend on the sites $l$
and $l'$ since the wavefunctions $f_l$ are all the
same.
  
In order to obtain the excitation frequencies, one has to linearize
Eqs.(\ref{jos1},\ref{jos2}) around equilibrium: in Eq.(\ref{jos1}) it
is enough to take the value of $\delta^{l,l'}$ at equilibrium; instead
in Eq.(\ref{jos2}) one has to expand $\delta_{\mu}^{l,l'}$ to first
order in the density fluctuations $\Delta n_l$
  
\begin{eqnarray}  
\delta_{\mu}^{l,l'} \approx \delta_{\mu} +  
\frac{\partial \delta_{\mu}^{l,l'}}{\partial n_l} \Delta n_l+  
\frac{\partial \delta_{\mu}^{l,l'}}{\partial n_{l'}} \Delta n_{l'}.  
\end{eqnarray}  
This procedure allows us to recover exactly Eq.(\ref{approx_bog_t-b}).  
Instead, if one sets $\sqrt{n_l n_{l'}}=n_l$ from the beginning one
obtains result (\ref{bog-tb-larges}) for the excitation spectrum.
This proves that the first term in the brackets of Eq.(\ref{approx_bog_t-b}) has its
physical origin in the quantum pressure, because it arises from the difference
in population $n_l-n_{l'}$ between neighbouring sites.

Note that in the form (\ref{jos1},\ref{jos2}), the equations of motion differ from 
commonly used Josephson equations in that the quantities $\delta^{l,l'}$ and $\delta_{\mu}^{l,l'}$ 
can be time-dependent. 
In the usual treatment, these quantities are calculated at equilibrium and one approximates 
$\delta=\delta_{\mu}$. 
The resulting simplified equations are equivalent to the discrete nonlinear Schr\"odinger equation 
used for example in \cite{trombettoni1} to investigate nonlinear phenomena like solitons and
breathers. 
An approach equivalent to the Josephson formalism presented in this section, based on 
the ansatz (\ref{tb_ansatz}), has been developed independently in \cite{trombettoni}, 
where the approximations involved are discussed in detail.

For a system confined in the radial direction, containing $N$ atoms per site, it is also 
convenient to introduce the Josephson energy $E_J=N\delta$ and the charging energy 
$E_C=2\partial\mu/\partial N=2\kappa^{-1}/N$ which play an important role in the physics 
of Josephson oscillations.

\section{Quantum fluctuations and depletion of the condensate}    
\label{fluct_depl}    

The presence of the optical potential may introduce phase fluctuations
which reduce the degree of coherence of the sample. This effect is known
to yield spectacular consequences in 3D optical lattices, giving rise to
a transition from the superfluid to the Mott insulator phases. Also
in the presence of a 1D optical lattice one can predict interesting
effects. 
First of all, the quantum depletion of the condensate increases 
as a consequence of the increase of the effective  coupling
constant (\ref{gtilde}) and of the effective mass. 
Eventually, if the tunneling rate becomes very small, 
the system reduces to a 1D chain of Josephson
junctions with a modification of the behaviour of long range order affecting 
the phase coherence of the system.

Let us consider the problem in a 3D box, with the optical
lattice oriented along the $z$-direction (Note that the quantum depletion 
of the condensate has been calculated in \cite{oosten,rey} for different geometries).  
The quantum numbers of the elementary excitations are
the band index $j$ and the quasi-momentum $q$ along the $z$ direction
and the momenta $p_x$ and $p_y$ in the transverse directions.  The
quantum depletion of the condensate can be calculated using the
Bogoliubov result

\begin{eqnarray}
{\Delta N_{\rm tot} \over N_{\rm tot}} = \!\!{1\over N_{\rm tot}}\!\! 
\sum_j \!\sum_{q,p_x,p_y} 
\int_{-d/2}^{d/2}\!\!\!\!\!\!dz\!\! \int \!\!dx\!\! \int\!\!dy \;|v_{j,q,p_x,p_y}({\bf r})|^2,
\label{depletion1}
\end{eqnarray}
where $N_{\rm tot}$ denotes the total number of atoms, $\Delta N_{\rm tot}$ is the number of 
non-condensed particles and 
we sum over all bands $j$, over the quasi-momenta $q$ in the
first Brillouin zone and the momenta of elementary excitations in the
transverse directions $p_x,p_y$ allowed by the periodic boundary conditions.
Eq. (\ref{depletion1})
describes correctly the depletion when Bogoliubov theory
is applicable.

In the thermodynamic limit, the depletion can be calculated replacing the sum with an integral in
Eq.(\ref{depletion1}).  In the uniform case, the main
contribution to the depletion is given by quasi-particles with 
$q^2+p_x^2+p_y^2 \approx (mc)^2$. Yet the convergence
is very slow and the integral is saturated by momenta much higher than 
$mc$ \cite{muntsa}, where the dispersion exhibits the quadratic 
$p^2/2m$ behaviour.
This implies that in the presence of the lattice 
it is possible to calculate the depletion as in the
uniform case, provided all the quasimomenta relevant for the calculation of the depletion 
lie within the first Brillouin zone.
This zone should include a region beyond the phononic regime where the dispersion goes like 
$q^2/2m^*$. 
This condition is satisfied if the inequality $m^* c \ll q_B$ (or equivalently $\kappa^{-1} \ll \delta$),
corresponding to weak interactions and relatively low values of $s$, is fulfilled.
Under this 
condition, we can replace $q^2/2m \to q^2/2m^*$ and $g \to {\tilde
g}$ and the quantum depletion takes the generalized Bogoliubov form
\begin{equation}
{\Delta N_{\rm tot}\over N_{\rm tot}}
={8\over 3}{1\over \pi^{1/2}}\sqrt{m^*\over m}(\tilde{a}^3n)^{1/2}
\, ,
\label{depletion3D}
\end{equation}
where we have defined $\tilde{a}$ through the relation 
$\tilde{g}=4\pi\hbar^2 \tilde{a}/m$. In this regime, the
depletion will not be quantitatively very different from the
corresponding one in the absence of the lattice.
The situation becomes more interesting for larger optical potential
depth, where the lattice is expected to affect the coherence
properties of the system.

In the regime of deep optical lattices, one can neglect contributions
to the depletion from higher bands, because high energy excitations are
particle-like. We are then allowed to restrict the sum in
(\ref{depletion1}) to $j=1$.
In the tight binding limit, one can easily generalize expressions
(\ref{t-b-u},\ref{t-b-v}) for the Bogoliubov amplitudes in the lowest band to account for
transverse excitations
\begin{eqnarray}  
\label{t-b-u-3d}  
u_{q,p_x,p_y}({\bf r})&=& {e^{i(p_x x+p_y y)/\hbar}\over L} U_{q,p_x,p_y}\! 
\sum_l e^{i q l d / \hbar} f(z\!-\!ld),\\  
v_{q,p_x,p_y}({\bf r})&=& {e^{i(p_x x+p_y y)/\hbar} \over L} V_{q,p_x,p_y}\!
\sum_l e^{i q l d/\hbar} f(z\!-\!ld),
\label{t-b-v-3d}  
\end{eqnarray}  
where $f(z)$ is the same function as in (\ref{t-b-varphi}) and $L$ is the transverse size of the system. 
Neglecting for simplicity contributions arising from
$n{\partial\delta/\partial n}$, Eq.(\ref{approx_bog_t-b}) can be
generalized in a straightforward way to
\begin{eqnarray}    
\label{approx_bog_t-b-3d}    
\hbar \omega_q\!&\approx&\! 
\sqrt{ \varepsilon_0(p_{\perp},q) (\varepsilon_0(p_{\perp},q)+ 2 \kappa^{-1} )},
\end{eqnarray}    
where 
$\varepsilon_0(p_{\perp},q) = {p_{\perp}^2/ 2m} + 2\delta\sin^2(qd/2\hbar)$ and $p_{\perp}^2=p_x^2+p_y^2$.
For the amplitudes $U_{q,p_x,p_y}$ and $V_{q,p_x,p_y}$ we find the result
\begin{eqnarray}    
U_{q,p_x,p_y}\,, V_{q,p_x,p_y}
&=&
\frac{ \varepsilon_0 \pm \hbar \omega}
{2 \sqrt{\hbar \omega \varepsilon_0}} ,
\label{bog-V-3d}
\end{eqnarray}
satisfying the normalization condition 
$\int_{-d/2}^{d/2}\! dz \int\!\!dx\!\!\int\!\!dy
\left[|u_{q,p_x,p_y}|^2-|v_{q,p_x,p_y}|^2\right]=1$.

We replace again the sum (\ref{depletion1}) by an integral.  This
corresponds to considering the thermodynamic limit in all the 3 directions.
Inserting Eqs.(\ref{t-b-v-3d},\ref{bog-V-3d}), the calculation can be performed analytically and gives
\begin{equation}
{\Delta N_{\rm tot}\over N_{\rm tot}}
=
2 \:\:{\tilde{a}\over d} \:\:G\left({\delta\over\tilde{g}n}\right)\,,
\label{resultdepletion}
\end{equation}
where $G(b)={1/ 2}-{\sqrt{b}/\pi}+{b/2}-{\rm arctan}(\sqrt{b})(1+b)/\pi$ and, for simplicity 
we have used the approximation $\kappa^{-1}=\tilde{g}n$
for the compressibility. 
Since in the tight binding regime the ratio 
${\delta/\tilde{g}n}$ is usually small and becomes smaller and smaller 
with increasing lattice depth, the 
result (\ref{resultdepletion}) converges to 
\begin{equation}
{\Delta N_{\rm tot}\over N_{\rm tot}}
=
{\tilde{a}\over d}\,.
\label{2D}
\end{equation}
Result (\ref{2D}) coincides with 
the 2D depletion of a disc of axial extension $d$ and
scattering length $\tilde{a}$, where the axial confinement is so
strong that the motion is frozen along the $z$-direction.  
In this limit the actual 3D system is described as a series of separated 2D discs. 
It is 
interesting to note that the depletion remains finite even if the tunneling
parameter $\delta \to 0$.  The
reason is that before taking the limit $\delta\to 0$ we have
taken the continuum limit in the radial direction, which
forces the system to be coherent. 
In this case the ratio $E_J/E_C=N\kappa\delta/2$ between the 
Josephson energy and the charging energy (see end of the last section) is large and 
hence coherence is maintained across the whole sample.
A different result would be obtained by fixing $N$ and considering the limit $\delta\rightarrow 0$.
We point out that the dependence on the interaction strength is stronger
in the 2D case than in the 3D case, since the depletion scales
like ${\tilde a}$ (2D) rather than ${\tilde a}^{3/2}$ (3D).

From Eq.(\ref{resultdepletion}), one recovers the 3D result (\ref{depletion3D})
in the limit $\delta/{\tilde g}n \to \infty$. This limit 
is by the way only of academic interest, since in the
tight binding limit, where (\ref{resultdepletion}) was derived, the ratio
$\delta/{\tilde g}n $ becomes large only if interactions are vanishingly
small 
(For example, for $gn=0.02 E_R$ and $s=10$ one finds $\delta/\tilde{g}n\approx 1$.).

Under the assumption made in the continuum approximation,
the depletion is always very small and is upper-bounded by the quantity ${\tilde a}/d$.
If the continuum approximation
in the radial direction is not applicable and it is crucial to take into
account the discretization of the sum over the quantum numbers $p_x$ and 
$p_y$ in Eq.(\ref{depletion1}), the results are different. This is the case if the number of
particles in each well is sufficiently small, or if the longitudinal size
of the system, fixed by the number of wells $N_w$, is sufficiently large.
The limiting case takes place when the contribution arising from the
term with $p_x=p_y=0$ is the dominant one in Eq.(\ref{depletion1}). In
this case the system exhibits typical 1D features and one finds the result
\begin{eqnarray}
{ \Delta N_{\rm tot} \over N_{\rm tot}} = 
\nu \ln \left( \frac{4 N_w}{\pi} \right),
\label{1d}
\end{eqnarray}
where 
\begin{eqnarray}
\nu = \frac{m^* c d }{ 2 \pi \hbar N}\,.
\label{nu}
\end{eqnarray}
Here, $N$ is the number of particles per well. 
The dependence on the interaction strength in
the 1D case is stronger than in the 2D and 3D cases, since 
$c=\sqrt{\tilde{g}n/m^*}$ and hence the
depletion scales like ${\tilde a}^{1/2}$. 
This should be compared with the ${\tilde a}$ and $ {\tilde a}^{3/2}$ dependence in 2D and 3D respectively.

The transition to the 1D character of the fluctuations can be
identified by the condition
\begin{eqnarray}
\frac{\tilde a}{d} \approx  \nu \ln \left( \frac{4 N_w}{\pi} \right) ,
\label{2d1d}
\end{eqnarray}
which, for $gn=0.2 E_R$, $N_w=200$ and $N=500$, is predicted to occur 
around $s=30$ where the left and right side of the inequality become equal to 
$\sim 4\%$.

\pagebreak
Result (\ref{1d},\ref{nu}) is strictly linked to the coherence theory of 1D systems,
where the off-diagonal 1-body density exhibits the power law decay 

\begin{eqnarray}
n^{(1)}(|{\bf r}-{\bf r}'|) \to |{\bf r}-{\bf r}'|^{-\nu}
\end{eqnarray}
at large distances.  If the exponent $\nu$ is much smaller than $1$,
the coherence survives at large distances and the application of
Bogoliubov theory is justified. For a superfluid, the value of $\nu$
is fixed by the hydrodynamic fluctuations of the phase and is given,
at $T=0$, by the expression (\ref{nu}) \cite{book,nolattice}.  In
terms of the Josephson parameters (see section \ref{josephson}) 
one can also
write $\nu = \sqrt{E_C/8\pi^2 E_J}$.  One can easily check that,
unless $N$ is of the order of unity or $m^*$ is extremely large, the
value of $\nu$ always remains very small.  When the exponent of the
power law takes the value $\nu = 0.14$, corresponding to $E_J=1.62
E_C$, the 1D system is expected to exhibit the Bradley-Doniach phase
transition to an insulating phase where the 1-body density matrix
decays exponentially \cite{bradley}.
Note however that before this transition is reached the depletion 
(\ref{1d}) becomes large and hence Bogoliubov theory is no longer applicable.

To give an example, we set $gn=0.2 E_R, N=200$ and $N_w=500$ 
describing a setting similar to the experiment of \cite{cataliotti}. 
Bogoliubov theory predicts a depletion of 
$\approx 0.6\%$ in the absence of the lattice ($s=0$). 
At a lattice depth of $s=10$ the evaluation of Eq.(\ref{depletion1}), using the tight binding 
results (\ref{t-b-v-3d},\ref{bog-V-3d}), and keeping the sum discrete yields 
a depletion of $\approx 1.7\%$. 
On the other hand, Eq.(\ref{resultdepletion}), obtained by replacing the sum in Eq.(\ref{depletion1}) by an integral, yields a depletion of $\approx 2\%$, in reasonable agreement with the full result  $\approx 1.7\%$.
The 2D formular (\ref{2D}) instead yields $\approx 2.9\%$ depletion, revealing that the system is not yet fully governed by 2D fluctuations. 
With the same choice of parameters, 
the power law exponent (\ref{nu}) has the value $\nu=0.001$ 
and the 1D depletion (\ref{1d}) is predicted to be $\approx 0.6\%$, 
significantly smaller than the full value 
$\approx 1.7\%$.  
This reveals that the sum (\ref{depletion1}) is not exhausted by the terms with 
$p_x=p_y=0$. 
In conclusion, one finds that for this particular setting, 
the character of fluctuations is intermediate between 3D and 2D, and still far from 1D. 
In particular, from the above estimates it emerges that in order to reach the conditions for 
observing the Bradley-Doniach transition one should work at much larger values of $s$.

\section{ Applications to harmonically trapped condensates}    
\label{appl_to_ho}    

The results obtained in the previous sections can be used to describe
harmonically trapped condensates in the presence of an optical
lattice: If the trapped condensate is
well described by the TF-approximation in the absence of the lattice 
and if the axial size of the
condensate is much larger than the interwell separation $d$, then one
can generalize the local density approximation (LDA) to describe 
harmonically trapped condensates in a lattice.  
Basically, the idea is to introduce the average density 
\begin{equation}  
n_l(r_{\perp})={1\over d}\int_{ld-d/2}^{ld+d/2} \!\!n(r_{\perp},z)\,dz\; \,,  
\label{profile_lda}  
\end{equation}  
where $n(r_{\perp},z)$ is the microscopic density and $l$ is the index of 
the lattice sites.  
Within the LDA, the chemical potential is given by
\begin{equation}  
\mu_l=\mu_{\rm opt}(n_l(r_{\perp}))+  
{m\over 2}(\omega_z^2l^2d^2+\omega_{\perp}^2r_{\perp}^2)\,,  
\label{mu_lda}  
\end{equation}  
where $\mu_{\rm opt}(n_l(r_{\perp}))$ is the chemical potential calculated at
the average density $n_l(r_{\perp})$ in the presence of the optical potential and
$\omega_z, \,\omega_{\perp}$ are the axial and transverse frequencies
of the harmonic trap respectively.  
Eq.(\ref{mu_lda}) fixes the radial density profile $n_l(r_{\perp})$ at the $l$-th site  
once the value of $\mu_l$ or, equivalently, the number of atoms 
\begin{equation}  
N_l=2\pi d \int_0^{R_{l}} r_{\perp}dr_{\perp}n_l(r_{\perp})\,,  
\label{normlr0}  
\end{equation}  
occupying the $l$-th well is known. 
In Eq.(\ref{normlr0}), 
$R_{l}$ is the radial size of the condensate at the $l$-th site, fixed by the value of 
$r_{\perp}$ where the density $n_l(r_{\perp})$ vanishes. 
This procedure avoids the full calculation of the microscopic density $n(r_{\perp},z)$.   

When equilibrium is established across the whole sample we have
$\mu_l=\mu$ for all $l$.  Making use of this fact and employing that
$\sum_l N_l=N_{\rm tot}$ we can find the dependence of $\mu$ on the total
number of particles $N_{\rm tot}$.  
This procedure also yields the well
occupation numbers $N_l$ and the number of sites occupied at
equilibrium.
In the simple case in which the chemical potential exhibits a linear dependence on density 
$\mu_{\rm opt}=\tilde{g}n+const.$ one obtains for the radial density profile \cite{pedri} 
\begin{equation}  
n_l(r_{\perp})={1\over\tilde{g}}\left(\mu-{m\over2}\omega_z^2l^2d^2-{m\over 2}\omega_{\perp}^2r_{\perp}^2\right)\,,
\end{equation}  
where the chemical potential, apart from a constant, is given by 
$\mu=\hbar\bar{\omega}(15 N_{\rm tot}a\,\tilde{g}/a_{ho}g)^{2/5}/2$ 
with $\bar{\omega}=(\omega_x\omega_y\omega_z)^{1/3}$, $a_{ho}=\sqrt{\hbar/m\bar{\omega}}$. 
The well occupation numbers and transverse radii are given by
\begin{eqnarray}  
N_l&=&N_0\left(1-l^2/l_m^2\right)^2\,,\\
R_l&=&R_0\left(1-l^2/l_m^2\right)^{1/2}\,,
\end{eqnarray}
where $l_m=\sqrt{2\mu/m\omega_z^2d^2}$ fixes the number $2 l_m+1$ of occupied sites, $N_0=15 N/16 l_m$ and $R_0=\sqrt{2\mu/m\omega_{\perp}^2}$. 
The increase of $\mu$ due to the optical lattice ($\tilde{g}>g$) implies an increase of the radii $R_l$. This effect has been observed in the experiment of \cite{morsch1}.

The LDA-based approach not only permits to calculate equilibrium properties, but also dynamic features of macroscopic type. 
To this purpose one can generalize the hydrodynamic equations of superfluids by taking into account the effects of the lattice. 
Also in this case
one can use the concept of the average density profile $n_l(r_{\perp})$ 
as defined in Eq.(\ref{profile_lda}).
Furthermore, one can replace the
discrete index $l$ by the continous variable $z=ld$.  In this way, one
can define a smoothed macroscopic average density profile
$n_{M}(r_{\perp},z)$.  The dynamics we are interested in then consists
of the evolution of the macroscopic density $n_M$ and of the 
macroscopic superfluid velocity field ${\bf v}$ whose 
component along the direction of the lattice is defined by the average
\begin{eqnarray}   
v_z&=&{1\over D}\int_{-D/2}^{D/2}\!\!dz\left[{\hbar\over m}\partial_z S\right]
\nonumber\\
&=&{\hbar\over m}{1\over D}\left(S(D/2)-S(-D/2)\right)\,.
\label{v_s}
\end{eqnarray}
Here $S$ is the phase of the order parameter and $D$ is a length 
longer than $d$, but small compared to the size of the system as well as 
to the wavelength of the collective oscillations. 
On this length scale, the effect of $V_{\rm ho}$ can be neglected and 
all the macroscopic variables are constant. 
Hence the phase difference in (\ref{v_s}) can be calculated using 
Bloch states with quasimomentum $k$ (see Eq.(\ref{bloch-ansatz})) and  
one finds the result $v_z=k/m$ \cite{notevz}. 
Using this result and the fact that the energy change per particle 
due to the presence of a small current is 
$k^2/2m^*=m^2v_z^2/2m^*$, the use of the LDA 
yields the result 

\begin{eqnarray}   
E=\int d{\bf r}   
\left[  
{m\over 2}n_M v_x^2  
+  
{m\over 2}n_M{v_y^2}  
+  
{m\over 2}{m\over m^*}n_M{v_z^2}\right.\nonumber\\  
+\left.  
e(n_M)  
+  
n_M V_{\rm ho}  
\right]\,,  
\label{lda-energy-1}  
\end{eqnarray}  
for the total energy of the system. 
Here $e(n_M)$ is the equilibrium energy per unit volume calculated at the average density   
$n_M$ in the absence of the harmonic trap and both $n_M$ and $v_z$ are now functions of $r_{\perp},\,z$ and $t$.

Starting from the functional (\ref{lda-energy-1}) we can derive the hydrodynamic equations 
\begin{eqnarray}  
{\partial\over \partial t}n_{M}  
&+&  
\partial_x({v_x}n_{M})  
+  
\partial_y({v_y}n_{M})  
+  
\partial_z\left({m\over m^*}{v_z}n_{M}\right)  
=0\label{hydro1}\\  
m{\partial\over \partial t}{\bf v}  
&+&  
\nabla\left(V_{\rm ho}  
+\mu_{\rm opt}(n_{M})\right.\nonumber\\  
&+&\left.{m\over 2}v_x^2+{m\over 2}v_y^2  
+{\partial\over\partial n_M}
\left(\left({m\over m^*}n_M\right){m\over 2}v_z^2\right)\right)  
=0\,,  
\label{hydro2}  
\end{eqnarray}  
where $V_{\rm ho}= {m}(\omega_z^2z^2+\omega_{\perp}^2r_{\perp}^2)/2$
is the harmonic potential and where we have used the relationship 
$\mu_{\rm opt}=\partial e(n_M)/\partial n_M$. 
Eqs.(\ref{hydro1},\ref{hydro2}) generalize the hydrodynamic equations derived in \cite{meret} 
to situations in which the effective mass is density dependent and the chemical potential $\mu_{\rm opt}$ has a  nonlinear density dependence (see section (\ref{section2})). 
They can be further generalized to account for 
larger condensate velocities \cite{machholm}.

The hydrodynamic equations serve to determine the frequencies of
collective oscillations associated with a density dynamics of the form
$n_M({\bf r},t)=\bar{n}_M({\bf r})+e^{-i\omega t}\delta n({\bf r})$.
Here, $\delta n({\bf r})$ denotes a small deviation from the
equilibrium average density $\bar{n}_M({\bf r})$ that is associated
with a small velocity field ${\bf v}$.  Hence, it is appropriate to
linearize Eqs.(\ref{hydro1},\ref{hydro2}) yielding
  
\begin{eqnarray}  
\omega^2\delta n &+&
\partial_{{\bf r}_{\perp}}\left[{\bar{n}_M\over m}\partial_{{\bf r}_{\perp}}
\left(\left.
{\partial\mu_{\rm opt}\over\partial n}\right|_{\bar{n}_M}\!\!\delta n\right)
\right]+
\nonumber\\  
&+&\partial_z\;\,\left[{\bar{n}_M\over m^*}\partial_z\;\,\,\left(\left.
{\partial\mu_{\rm opt}\over\partial  n}\right|_{\bar{n}_M}\!\!\
\delta n\right)\right]  
=0  \,.
\label{hydro3}  
\end{eqnarray}  
This second order equation involves the
derivative $\left.\partial\mu_{\rm opt}/\partial n\right|_{\bar{n}_M}$
directly related to the inverse compressibility $\kappa^{-1}$ (see Eq.(\ref{kappa})).  
In general, the solution of
Eq.(\ref{hydro3}) should be found numerically because 
the density dependence of $m^*$ and $\partial\mu_{\rm opt}/\partial n$ 
is not known in an analytic form. 
Yet, for small enough densities the density-dependence of
$m^*$ can be neglected and $\left.{\partial\mu_{\rm opt}/\partial
n}\right|_{\bar{n}_M}=\tilde{g}$ as was discussed in
Sect.(\ref{section2}).  In this case, the frequencies of the collective
oscillations can be calculated analytically \cite{meret}.
In particular, they 
can be obtained from the values in the absence of the lattice \cite{stringari} 
by simply rescaling the trapping frequency along $z$
  
\begin{equation}  
\omega_z\rightarrow \sqrt{m\over m^*}\omega_z\,.  
\end{equation}  
This result was obtained theoretically for the dipole oscillation in the tight binding regime 
in \cite{cataliotti} and has been confirmed experimentally both for the dipole \cite{cataliotti} 
and quadrupole oscillation \cite{fort}.

The LDA defined by (\ref{mu_lda}) can also by employed to study
Bogoliubov excitations occurring on a length scale much smaller than
the size of the system.  Under this condition, one can define a local
Bogoliubov band spectrum given by $\hbar\omega_j(q;n_M({\bf r}))$
where $n_M({\bf r})$ is the locally averaged density introduced
above.  The Bogoliubov band spectrum can be
probed by measuring the dynamic structure factor $S(p,\omega)$, related to
the linear response of the system to an external perturbation
transferring momentum $p$ and energy $\hbar\omega$.  In LDA, the
dynamic structure factor reads \cite{francesca}

\begin{equation}  
S_{LDA}(p,\omega)=\int \!\!d{\bf r} \,n_M({\bf r}) S(p,\omega;{\bf r})\,,
\label{struc-fac-lda} 
\end{equation}  
where $S(p,\omega;{\bf r})$ is the dynamic structure factor
(\ref{struc-fac}) calculated at density $n_M({\bf r})$ in the absence
of harmonic trapping.  As a result of the averaging implied by
Eq.(\ref{struc-fac-lda}), the dynamic structure factor of the system is not any more given by a
series of delta-functions as in the absence of the harmonic trap, but
instead consists of resonances of finite width.  The validity of the
LDA to describe the linear response of the trapped condensate in the
absence of a lattice has been confirmed by Bragg spectroscopy
experiments where the transferred momentum was larger than the inverse
of the system size and the duration of the Bragg pulse was small
compared to the inverse of the trapping frequencies \cite{bragg,dalfovo-davidson}.  
Experiments and calculations which investigate the regime beyond the validity of the
LDA have been recently carried out \cite{dalfovo-davidson,tozzo}.

In the absence of harmonic trapping, the solution of the hydrodynamic equations (\ref{hydro1}) and (\ref{hydro2}) permits to calculate the propagation of sound waves 
not only in condensates at rest, yielding result (\ref{sound_vel}) for the sound velocity, but also in condensates moving slowly with respect to the lattice. 
Let us start from stationary solution with macroscopic velocity $\bar{v}_z$, corresponding to quasimomentum $k=m\bar{v}_z$, with $k\ll q_B$, and let us consider small changes of the velocity field and of the density with respect to the stationary values: 
$v_z=\bar{v}_z+\delta v_z$, $n_M=\bar{n}_M+\delta n_M$. 
By linearizing Eqs.(\ref{hydro1}) and (\ref{hydro2}) and looking for solutions oscillating like $\sim e^{-i(qz-\omega t)}$, one finds at first order in $k$
\begin{eqnarray}
\omega=c|q|+q{k\over m^*_{\mu}}\,,
\label{omegaqk}
\end{eqnarray}
where $1/m_{\mu}^*=\partial(n/m^*)/\partial n$ and $c$ is the sound velocity in 
the condensate at rest (see Eq.(\ref{sound_vel})). 
The quantity $m_{\mu}^*$ gives the $k=0$ curvature of the lowest chemical potential band  
(see Eq.(\ref{mubands}) with $j=1$)  according to 
$\mu(k)=\mu(0)+k^2/2m_{\mu}^*$ for $k\rightarrow 0$.
Eq.(\ref{omegaqk}) generalizes the usual behaviour of the sound velocity in slowly moving frames 
to account for the presence of the optical lattice. 
The significance of $m_{\mu}^*$ for the excitation spectrum has been pointed 
out for any optical potential depth for small $q$ and any $k$ in \cite{machholm} and in 
the tight binding regime for any $q$ and any $k$ in \cite{trombettoni,menotti}.

\section{Summary}

We have studied Bloch-wave solutions of the Gross-Pitaevskii equation in the presence of a one-dimensional optical lattice. 
In particular, we have calculated the band structure of both 
stationary (``Bloch bands'') and time-dependent linearized (``Bogoliubov bands'') solutions. We have discussed these solutions 
for different choices of the lattice depth $s E_R$ and the two body interaction parameter $gn$. 
Special attention has been paid to the behaviour of the compressibility and of the effective mass. 
We have shown that the compressibility of the system is reduced by the presence of the lattice and that its inverse is approximately linear in the density for low enough $gn$ or high $s$. In these regimes, the compressibility and the chemical potential can be expressed in terms of an effective coupling constant $\tilde{g} > g$, which accounts for the squeezing of the condensate wavefunction in each well.
Concerning the effective mass we have found that two body interactions give rise to a significant 
density dependence which decreases its value with respect to the prediction for the non-interacting system. 
The compressibility and the effective mass permit to calculate the sound velocity whose value is found to decrease as a function of the lattice depth, reflecting the exponential increase of the effective mass. 
For the tight binding regime, we have complemented the numerical results by analytic expressions. 

Concerning the Bogoliubov bands, we have found that in a deep lattice the excitations in the lowest band acquire a strong quasi-particle character in the whole Brillouin zone, characterized by Bogoliubov amplitudes $u\sim v$.  
In the tight binding regime, analytic expressions for the lowest Bogliubov band and the corresponding Bogoliubov amplitudes have also been reported.

In section \ref{fluct_depl}, we have presented results for the quantum depletion of the condensate. 
In particular, we have found that in a deep lattice the quantum fluctuations of the condensate 
acquire 2D character, reflecting the transformation of the system into a series of two-dimensional discs. 
As a consequence, the quantum depletion increases, but remains small, 
provided the coherence between the discs is maintained. 
If the lattice depth is increased further, 1D quantum fluctuations become important. 
Estimates of the corresponding effects with realistic values of the parameters have been presented.

Finally, we have demonstrated the use of a local density approximation to study macroscopic static and dynamic properties of harmonically trapped systems in the presence of an optical lattice.  

In conclusion, Bose-Einstein condensates in optical lattices 
share important analogies with solid state systems. 
Differences arise due to the presence of two body interactions giving rise to important new features even in the coherent regime where most particles are in the condensate. 
The density dependence of the effective mass, as well as the distinction between Bloch, chemical potential and Bogoliubov bands are some examples discussed in this paper. 
Natural developments of the present work concern the study of the dynamics built on top of moving condensates where dynamic instabilities can be encountered \cite{niu,smerzi,machholm,trombettoni,menotti}. 
Another important direction is the study of nonlinear effects which might sizeably affect the propagation of sound in the presence of an optical lattice.

\section{Acknowledgments}

We would like to thank M.L.~Chiofalo, O.~Morsch and A.~Smerzi for useful discussions.  
This research is supported by the Mi\-ni\-ste\-ro dell'Istru\-zio\-ne,   
dell'Uni\-ver\-si\-t\`a e del\-la Ri\-cer\-ca (MIUR).


\begin{references}
\bibitem{notedeltamu}
Analogously, using Eq.(\ref{mu_l}), one finds the relation $\delta_{\mu}/E_R=2m/m^*_{\mu}\pi^2 $ where $1/m_{\mu}^*=\partial(n/m^*)/\partial n$ is the curvature of the band $\mu_{j=1}(k)$ at $k=0$ and $\delta_{\mu}$ has been defined in Eq.(\ref{t-b-mu}).
\bibitem{notevz} 
The superfluid velocity should not be confused with the group velocity (\ref{groupv}), which, for $k\to 0$, approaches the value $\partial\varepsilon/\partial k=k/m^*$.
\end{references}
\end{document}